# Smooth anti-reflective three-dimensional textures for liquid-phase crystallized silicon thin-film solar cells on glass

*David Eisenhauer,*[*,a] *Grit Köppel,*[a] *Klaus Jäger,*[a,b] *Duote Chen,*[b] *Oleksandra Shargaieva,*[a] *Bernd Rech,*[a] *and Christiane Becker*[a]

[a] Helmholtz-Zentrum Berlin für Materialien und Energie GmbH, Kekuléstr. 5, 12489 Berlin, Germany

[b] Zuse-Institut Berlin, Takustr. 7, 14195 Berlin, Germany

**Abstract**

Recently, liquid-phase crystallization of thin silicon films has emerged as a promising candidate for thin-film photovoltaics. On 10 μm thin absorbers, wafer-equivalent morphologies and open-circuit voltages were reached, leading to a record efficiency of 12.1%. However, short-circuit current densities are still limited, mainly due to optical losses at the glass-silicon interface. While nano-structures at this interface have been shown to efficiently reduce reflection, up to now these textures caused an increase in recombination. Therefore, optical gains were mitigated due to electronic losses. Here, the SMooth Anti-Reflective Three-dimensional (SMART) texture is introduced in order to overcome this trade-off. By smoothing nanoimprinted $SiO_x$ nano-pillar arrays with spin-coated $TiO_x$ layers, light-trapping properties of laser-crystallized silicon solar cells could significantly be improved as successfully shown in three-dimensional simulations and in experiment. At the same time, the smooth surface morphology of SMART textures allows preserving electronic material quality equivalent to that of planar reference samples and reaching $V_{oc}$ values above 640 mV in 8 μm thin liquid-phase crystallized silicon solar cells. Furthermore, the short-circuit current density $j_{sc}$ could be increased from 21.0 mA cm$^{-2}$ for planar reference cells with already optimized anti-reflective interlayer stacks to 23.3 mA cm$^{-2}$ on SMART textures, corresponding to a relative increase of 11%.

# 1. Introduction

Liquid-phase crystallization of 5 - 40 μm thin silicon films directly on a glass substrate is presently a promising technology endorsing the general trend towards reduced absorber thicknesses in silicon photovoltaics. This technique allows avoiding current challenges of silicon wafers, namely high material losses and handling issues particularly arising at very low wafer thicknesses. By scanning a line-shaped energy source, e.g. a laser beam, across silicon films on glass large-grained polycrystalline material is formed [1–4]. It has been shown that an excellent material quality equivalent to that of multi-crystalline silicon wafers can be obtained using this technology, leading to record open-circuit voltages ($V_{oc}$) of 656 mV [5] and 12.1% solar cell efficiency [6]. The short-circuit current density ($J_{sc}$) of this record cell is limited to 28.9 mA cm$^{-2}$ due to reflection losses, mainly at the glass-silicon interface [6].

One possibility to reduce these losses is nano- or micro-structuring of the glass-silicon interface, either by directly texturing the glass [7, 8] or by nanoimprint lithography using high-temperature stable sol-gel films [9, 10]. While these measures efficiently enhance light-incoupling into the silicon absorber in superstrate devices, the textured interfaces cause lower material quality and higher surface recombination velocities [11].

Similar challenges are known for other silicon thin-film technologies, e.g. nano-crystalline silicon (nc-Si) thin-film solar cells [12–15]. One successful approach for nc-Si solar cells in substrate configuration is the implementation of flat light scattering substrates (FLiSS) as light-trapping structures [16–19]. In the FLiSS approach, a periodically patterned [16] or randomly textured [17, 18] ZnO layer is covered by an amorphous silicon layer and subsequently polished until the tips of the ZnO pattern are bared. Using the FLiSS substrate in nc-Si solar cells lead to an equivalent material quality of the absorber compared to planar devices, and a relative increase in efficiency of 10% was observed [18].

Another technique, which was developed for amorphous silicon thin-film solar cells in superstrate configuration, uses imprinting of random nano- and micro-pyramids in a hydrogen silsequioxane layer [20]. These pyramids are planarized by spin-coating a ZnO nanoparticle solution, thereby achieving material quality equivalent to that of planar devices while increasing the short-circuit current density, leading to an 18% relative efficiency increase.

As shown in Ref. [10], electronic material quality of liquid-phase crystallized silicon solar cells could be increased by smoothing hexagonal nano-pillar arrays to a sinusoidal texture. However, electronic material quality was still reduced as compared to a planar reference stack.

In this contribution, we present an alternative method for producing high-quality laser-crystallized silicon solar cells on nanostructured substrates, the SMooth Anti-Reflective Three-dimensional (SMART) texture. The SMART texture is produced by combining nanoimprint lithography of hexagonal nano-pillar arrays with spin-coating of titanium oxide layers. The spin-coating leads to a preferential filling of the voids between the hexagonal nano-pillars, resulting in a smooth surface morphology without edges and steep flanks. We present the compatibility of the SMART texture with the silicon laser crystallization and solar cell preparation process. Optical, electronic material quality and optoelectronic characterization of silicon thin-film solar cells on SMART textures reveal that these solar cells outperform reference cells on optimized planar interlayer systems.

## 2. Experimental and numerical methods
### 2.1 SMART texture

As superstrates 1.1 mm thick Corning Eagle XG glasses with a size of 5 x 5 cm² are used. After a cleaning step with a 5% alkaline Mucasol© aqueous solution, 250 nm thick SiOx is sputtered onto the superstrates, serving as a diffusion barrier to glass impurities during liquid-phase crystallization. Figure 1 illustrates the production process for SMART textures. Hexagonal pillar arrays with a period p of 750 nm and pillar heights h of 50 nm are replicated in a high-temperature stable, UV curable sol-gel resist based on silicon alcoxides [21] (step 1) using nanoimprint lithography [22]. Further details about the used nanoimprint process can be found in Ref. [23]. In order to smooth the surface of the superstrates, a titanium oxide precursor solution consisting of a mildly acidic solution of titanium isopropoxide in anhydrous ethanol is cast on the superstrate and spun with 2000 rpm for 30 s. The spin-coating results in a preferential filling of the voids between the silicon oxide pillars, thus reducing the surface roughness significantly (step 2). Through thermal curing for 30 minutes at 150°C and 30 minutes at 500°C, the solvents evaporate and a compact titanium oxide layer is formed.24 Finally, a 10 nm thin silicon oxide layer is sputtered onto the stack, serving as a passivation layer at the interface with the silicon absorber (step 3).

Figure 2a shows atomic force microscope images of the hexagonal nano-pillar array (step 1) and the SMART texture after spin-coating (step 3). For comparison, the height scaling was set constant in the measurements. It is clearly seen that the surface roughness of the substrate is greatly reduced by spin-coating the titanium oxide. This is also observed in the scanning electron microscope image in Fig. 2b. The $TiO_x$ layer (colored in green) preferably fills the voids between the $SiO_x$ nano-pillars (colored in blue). Specifically, edges and steep flanks of the texture – which are detrimental to silicon material quality after liquid-phase crystallization [10, 11] – are flattened out to a very smooth surface morphology at the interface to the crystalline silicon (c-Si, grey).

### 2.2 Absorber and solar cell preparation

The silicon absorber is deposited onto the SMART superstrates and on planar reference superstrates. The reference sample has an interlayer stack of 250 nm silicon oxide, 70 nm silicon nitride and 10 nm silicon oxide, which was previously found to have optimal anti-reflective properties [2, 4], and is processed in parallel to the SMART superstrates. The 8 μm thick silicon absorbers are deposited by electron beam evaporation at a substrate temperature of $T = 600°C$ [24].

Liquid-phase crystallization of the silicon is performed using a line-shaped laser with a 1/e² size of $30 \times 0.3$ mm$^2$ and a wavelength of 808 nm. The laser beam is scanned across the samples with a scanning speed of 3 mm/s, and melts the nano-crystalline silicon layer. Upon solidifying, large grains of up to centimeters in length and millimeters in width are formed [1, 5].

Solar cells with n-type doping concentration of about $1 \times 10^{17}$ cm$^{-3}$ are prepared on both substrate types as described in Refs. [4] and [5], therein denoted as test structure and test cells, respectively. Because of their high series resistance, these test devices have only low fill factors. Nevertheless, all vital solar cell parameters can be measured with this contacting scheme, which allows quick and simple processing [5].

## 2.3 Characterization

Optical characterization is conducted with a Perkin Elmer Lambda 1050 spectrophotometer equipped with an integrating sphere. The measured reflection $R(\lambda)$ and transmission $T(\lambda)$ spectra give the absorption $A(\lambda) = 1 - R(\lambda) - T(\lambda)$. The absorption can then be used to calculate the maximum achievable short-circuit current density ($j_{sc,max}$) with the equation

$$j_{sc,max} = e \int_{300\,nm}^{1100\,nm} A(\lambda)\Phi(\lambda)d\lambda, \qquad \text{Equation 1}$$

where $e$ is the elementary charge and $\Phi(\lambda)$ the spectral photon flux corresponding to the AM1.5G solar radiation spectrum.

Open-circuit voltages ($V_{oc}$) are determined using a WCT-100 photo conductance lifetime tool by Sinton Instruments. The setup additionally allows to calculate the so-called pseudo fill factor *pseudoFF*, assuming no series resistance in the device.

External quantum efficiency (EQE) is measured on a custom-made setup featuring a probe beam of 3 x 2 mm$^2$ using lock-in technique. While no bias voltage is applied during measurements, bias light from a halogen lamp is used, imitating the AM1.5G spectrum. Short-circuit current densities are also calculated from the EQE according to

$$j_{sc,EQE} = \int_{300\,nm}^{1100\,nm} EQE(\lambda)\Phi(\lambda)d\lambda. \qquad \text{Equation 2}$$

As the EQE depends not only on absorption but also on electrical losses in the solar cell, the maximum achievable short-circuit current density $j_{sc,max}$ is the upper bound to $j_{sc,EQE}$.

Atomic force microscope images were measured using a Park Systems XE-70. Microscopy was performed with a Hitachi cold field emitter scanning electron microscope.

## 2.4 Three-dimensional optical simulations

Optical simulations are performed with the 3-dimensional finite-element method (FEM) solver JCMsuite [25]. FEM provides rigorous solutions to Maxwell's equations for a given structure that is discretized in the trial space. On every finite element, solutions to Maxwell's equations are approximated by a polynomial. Numerical accuracy is achieved by constraining the side lengths of the elements to the light wavelength in the corresponding material and adapt the polynomial degree during simulations.

Top and bottom of the computational domain are considered as infinite halfspaces, which is realized by using so-called perfectly matched layers (PML). By doing this, the computational cost of the simulations can be kept low [26]. On the sides of the computational domain periodic boundary conditions are applied.

The simulated structures consist of hexagonal nano-pillar arrays, as illustrated in the meshed unit cell in Fig. 3. The parameters characterizing the nanostructure are the period $p$, the height $h$ and the diameter $d$ of the nano-pillar, as indicated in Fig. 3. The area filling fraction *FF* of SiO$_x$ nano-pillars in the SMART texture is connected to $p$ and $d$ via

$$FF = \frac{\pi}{2\sqrt{3}}\left(\frac{d}{p}\right)^2 \qquad \text{Equation 3}$$

Light absorbed in the silicon layer or reaching the perfectly matched layer can be considered as coupled into the absorber. Hence it can be interpreted as $1 - R$ (reflectance), because the (parasitic) absorption in the interlayers can be neglected. Therefore, $1 - R$ represents a measurable parameter of light that cannot reach the back-side of the absorber layer due to a sufficiently short penetration depths in silicon. For our devices, this is up to a wavelength of about 600 nm.

## 3 Results
### 3.1 Simulation results
In order to identify a suitable experimental structure for the SMART texture, optical simulations of hexagonal nano-pillar arrays, as sketched in Fig. 3, were performed with varying periods ($p$), heights ($h$) and filling fraction ($FF$). The period $p$ was varied between 350 nm and 730 nm and the pillar height $h$ between 20 nm and 300 nm. The filling fraction $FF$ was set to 0.25, 0.5 and 0.75 by choosing the appropriate diameter $d$ of the pillar following Eq. 3.

Figure 4(a) shows the fraction of light coupled into the silicon absorber ($1 - R$) in the wavelength range 400 nm – 600 nm for three different filling fractions $FF$ as a function of nanostructure height $h$, for a fixed period $p$ of 730 nm.

Smaller $SiO_x$ nano-pillars diameters, i.e. a lower filling fraction $FF$, lead to decreased reflectance. This can be explained by interpreting the SMART texture as a single mixed medium between $SiO_x$ ($n = 1.5$) and $TiO_x$ ($n = 2.1$). Within this representation, a smaller filling fraction of $SiO_x$ nano-pillars corresponds to a higher effective refractive index. As the optimal refractive index between glass and silicon, given by the geometrical mean value, is around 2.4, higher effective refractive indices improve the anti-reflective properties at the interface. Filling fractions lower than 0.25 were not considered in simulation due to difficulties in their experimental production.

Therefore, $FF$ was set to 0.25 during simulations of various periods (Fig. 4(b)). The simulations predict an optimum height of the SMART texture of 40-50 nm, which is independent of the period. Comparing different periods of the nanostructure, it is seen that mean reflectance is independent of period down to 500 nm. For the smallest period of 350 nm (blue curve), $R$ is reduced. However, the difference in reflectance in the optimal thickness range between the smallest (blue) and largest (purple) period is only 2%. As it has been shown that the laser crystallization process yields better material quality for larger nanostructure periods [10], the structure used in experiment was chosen to have a period of 750 nm, height of 45 nm and a $FF$ of 0.3.

### 3.2 Optical properties
Optical characteristics of the prepared solar cells were measured in order to confirm their anti-reflective behavior at the glass-silicon interface. Figure 5 shows the reflectance spectra, which are plotted as *1-R* (solid curves in Fig. 5). *1-R* represents the proportion of light coupled into the absorber, where it is either absorbed or transmitted. It can be seen that the SMART texture (green) significantly reduces reflection at the glass-silicon interface in the wavelength regime up to about 800 nm in comparison to the optimized planar reference (black). Particularly, the minimum reflectance is only about 5% at a wavelength of 560 nm, of which 4% (absolute) are already reflected at the sun-facing planar air-glass interface. For longer wavelengths, the SMART texture and the planar reference stack reflect the same amount of light. The mean reflectance in the wavelength regime 400 nm – 600 nm, where the influence of the rear side of the 8 μm thick silicon absorber can be excluded, amounts to 16% (absolute) in the optimized planar reference stack and 9% (absolute) for the SMART texture.

The anti-reflective properties of the SMART texture are compared to a nano-pillar array (red, data from Ref. [10]) with the same period as the SMART texture, but higher height-to-period ratio. While in the SMART texture the spin-coated TiO$_x$ anti-reflective layer leads to a smooth interface, the pillar sample with a state-of-the-art SiN$_x$ anti-reflective coating exhibits a rough silicon interface. Despite its smooth surface morphology, the SMART texture shows similar anti-reflective properties. Only for wavelengths shorter than 450 nm, the reflection of the reference pillar structure is significantly lower compared to the SMART texture. Between 450 nm and 600 nm, the SMART texture has better anti-reflective properties compared to the pillar array. For the wavelength regime above 600 nm, light-trapping at the silicon-air interface in the double-side textured nano-pillar sample influences the optical properties and the samples are no longer comparable. Therefore, transmission spectra (dashed lines) are only shown for the SMART texture and corresponding planar reference solar cells. The measurements reveal less transmitted light for the SMART texture in the long wavelength regime. As the roughness of the back silicon surface in both cases is similar as confirmed by atomic force microscope images (not shown), the reduced transmittance is attributed to scattering of light at the SMART texture. The contrast in refractive indices of the mixed SiO$_x$/TiO$_x$ layer makes it optically rough, leading to refractive and therefore a longer light path through the absorber.

In summary, SMART textures enhance the absorption in the 8 μm thin Si layers, which is calculated from reflectance and transmission via $A = 1 - R - T$. This enhancement is due to both anti-reflective and scattering properties and leads to an increase of the maximum achievable short-circuit current density $j_{sc,max}$ from 24.7 mA cm$^{-2}$ in the optimized planar reference stack to 28.4 mA cm$^{-2}$ in the SMART texture, which is a relative increase of 13%.

### 3.3 Material quality

While there exists a wide variety of nanotextures that increase the optical absorption in crystalline silicon thin-film solar cell absorbers [7–10], the real challenge is to simultaneously obtain an excellent electronic material quality [11].

Here, the electronic material quality of the bulk laser-crystallized silicon absorbers on SMART textures and planar reference samples was evaluated by means of Suns-$V_{oc}$ measurements. Figure 6 shows the open-circuit voltage of the four best cells (solid squares) of the planar reference, the SMART texture and the nano-pillar array from Ref. [10]. Mean values are represented by open squares, boxes represent the standard error and whiskers show the standard deviation. The highest measured $V_{oc}$ is highlighted by a star. As the SMART texture smooths the edges and steep flanks of the nano-pillar array and thus the interface to the silicon absorber, the silicon material quality can significantly be increased, demonstrated by an open-circuit voltage gain of more than 200 mV.

$V_{oc}$ values for both the planar reference stack and the SMART texture samples are very high, with mean and maximum $V_{oc}$ values of 629 mV and 636 mV for the planar reference sample and 644 mV and 649 mV for the SMART texture cells, respectively. Therefore, the material quality of the bulk silicon absorber crystallized on SMART textures is at least equivalent to the planar reference stack. The increase of $V_{oc}$ values on SMART textures cannot solely be explained by the increased current (cf. section 3.4), which we estimate from a one-diode model to about $\Delta V_{oc,current} \approx 3\ mV$. Therefore, the increased $V_{oc}$ of silicon thin-film solar cells on SMART superstrates might be attributed to passivation properties of the TiO$_x$ layer, which are already known from literature [27–30]. However, further investigations are needed in order to elucidate this hypothesis, e.g. using current-voltage measurements [31].

## 3.4 Optoelectronic properties

As discussed in section 3.2, implementing the SMART texture at the glass-silicon interface leads to improved optical properties of the laser-crystallized silicon solar cells. Because extraction of electron-hole pairs is critical for solar cell performance, it is important that recombination remains at a low level, both in the bulk and at interfaces. In Fig. 7 measurements of the external quantum efficiency in superstrate configuration (EQE, solid lines) and reflectance (1-R, dashed lines) are shown. The planar reference and SMART texture were measured with an additional white-paint back reflector, while the pillar array exhibits a double-side texturing. As seen in the $V_{oc}$ measurements (cf. Fig. 6), the material quality of the laser-crystallized absorber on the pillar array is poor, resulting in a low EQE. This is not the case for the SMART texture (green), as its EQE is even higher compared to the parallel processed planar reference (black) for the whole wavelength range from 400 nm to 1100 nm. We attribute this to the smooth surface of the SMART texture compared to the nano-pillar array. Thus, it is confirmed that the increased light absorption in the laser-crystallized silicon absorber is not mitigated by an increased number of bulk material or interface defects. The reduced EQE in the short wavelength regime can be explained by parasitic absorption in the $TiO_x$ layer, cf. the absorption curve (dotted) for the superstrate with a SMART texture in Fig. 7. However, solar radiation is not very strong in this wavelength regime and current loss compared to the planar reference cell is only 0.06 mA cm$^{-2}$.

Short-circuit current densities are calculated from EQE measurements as described in section 2.3 and maximum values are summarized in Table 1 together with maximum achieved solar cell parameters obtained from Suns-$V_{oc}$ measurements (cf. section 2.2). By implementing the SMART texture at the glass-silicon interface, all relevant parameters for device performance could be significantly enhanced with respect to the nano-pillar array. Compared to a planar reference cell, the short-circuit current density of the solar cells on SMART texture was increased from 21.0 mA cm$^{-2}$ to 23.3 mA cm$^{-2}$. The open-circuit voltage is slightly higher for solar cells on SMART superstrates, which might be explained by an improved interface passivation. The *pseudoFF* remains at the level of planar reference devices. In conclusion, by combining these effects the potential power conversion efficiency is increased from 10.8% to 12.0%.

**Table 1**: Overview of solar cell parameters, obtained by EQE and SunsVoc measurements of liquid-phase crystallized silicon thin-film solar cells on the nano-pillar array, the SMART texture and planar reference cells.

| sample | max. $J_{sc}$ (mA cm$^{-2}$) | max. $V_{oc}$ (mV) | max. *pseudoFF* (%) | Max. *pseudo η* (%) |
|---|---|---|---|---|
| pillar, Köppel et al. | 12.6 | 492 | 69.1 | 4.3 |
| SMART texture | 23.3 | 649 | 79.3 | 12.0 |
| planar reference | 21.0 | 636 | 81.0 | 10.8 |

To further increase the short-circuit current density and thus power conversion efficiency in liquid-phase crystallized silicon thin-film solar cells, the SMART texture can be combined with additional light-management measures at the air-glass interface and silicon rear-side, as has been successfully demonstrated for planar devices.

## 4. Conclusion

We presented a novel nanostructure for increased light in-coupling in liquid-phase crystallized silicon thin-film solar cells: the SMooth Anti-Reflective Three-dimensional (SMART) texture. Nanoimprinted, high-temperature stable $SiO_x$ sol-gel nano-pillars were smoothed by spin-coating of $TiO_x$. Thereby, an optically rough nanostructure with a smooth surface morphology could successfully be prepared and integrated in the laser crystallization solar cell preparation process.

Three-dimensional optical simulations allowed to obtain suitable nanostructure parameters for experiment. Anti-reflective properties and process compatibility of the nanostructure was found to be optimal at a period $p = 730$ nm, a filling fraction $FF = 0.25$ and nanostructure height $h = 40 - 50$ nm.

Solar cells with a SMART texture were prepared using the parameters obtained from simulations. Despite its much smoother surface morphology, anti-reflective properties of the SMART texture were found to be equivalent to the more pronounced hexagonal nano-pillar array. Compared to optimized planar interlayer devices, an additional increase of scattering due to the contrast in refractive indices of the SMART texture was observed. Overall, this lead to maximum achievable short-circuit current densities in 8 μm thick liquid-phase crystallized silicon solar cells on glass of 28.4 mA cm$^{-2}$ for cells with a SMART texture and 24.7 mA cm$^{-2}$ for an optimized planar reference stack.

The pronounced texture of the nano-pillar array was previously found to cause poor material quality of laser-crystallized silicon solar cells, limiting the open-circuit voltage to below 500 mV. In contrast, implementing the SMART texture had no detrimental effects on material quality of the silicon absorber. Moreover, cells on the SMART texture exceed the planar reference cells in terms of maximum achieved $V_{oc}$ values, from 636 mV on planar reference cells to 649 mV on solar cells with a SMART texture.

Furthermore, measurements of the external quantum efficiency confirmed that the increased in-coupling of light is not mitigated by an increased number of defects at the glass-silicon interface. This lead to an enhancement in $j_{sc}$ from 21.0 mA cm$^{-2}$ in case of the planar reference to 23.3 mA cm$^{-2}$ on the SMART texture, an 11% increase on the SMART texture cell. The potential power conversion efficiency of solar cells could be increased from 10.8% to 12.0%.

Hence, the SMART texture allows overcoming the trade-off between optical and electronic properties by decoupling the anti-reflective properties from absorber structuring.

**Acknowledgement**

The authors gratefully acknowledge the support of M. Krüger, H. Rhein, E. Conrad, P. Sonntag and C. Klimm, Helmholtz-Zentrum Berlin für Materialien und Energie GmbH (HZB), for their help with solar cell preparation and SEM imaging. Simulation results were obtained at the Berlin Joint Lab for Optical Simulations for Energy Research (BerOSE) of Helmholtz-Zentrum Berlin für Materialien und Energie, Zuse Institute Berlin and Freie Universität Berlin. The German Ministry of Education and Research (BMBF) is acknowledged for funding the research activities of the Young Investigator Group Nano-SIPPE at HZB in the program NanoMatFutur (no. 03X5520).

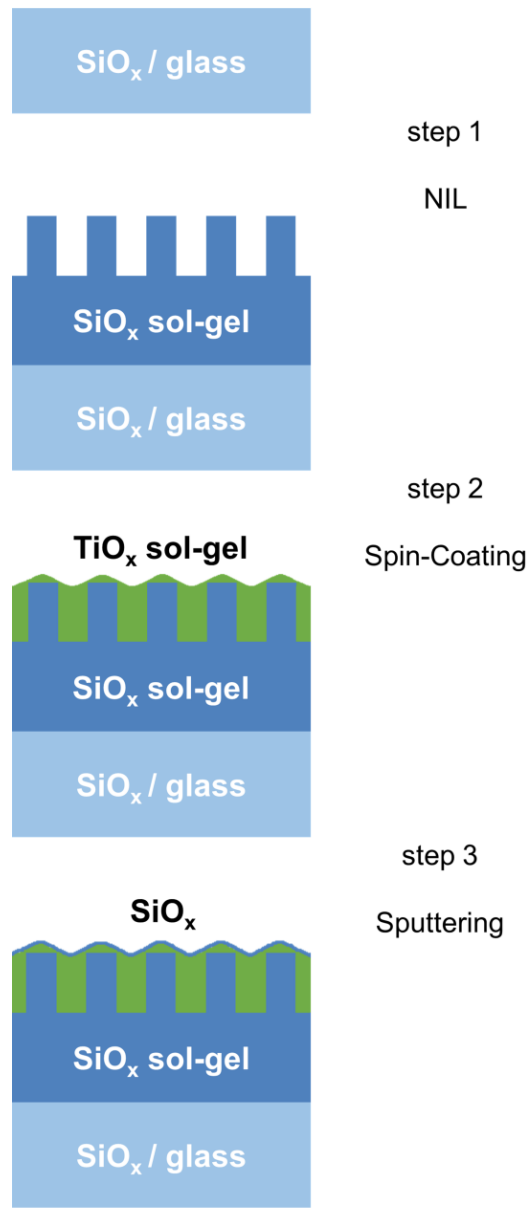

Fig. 1: Sketch of the schematic production process of superstrates with a SMART texture.

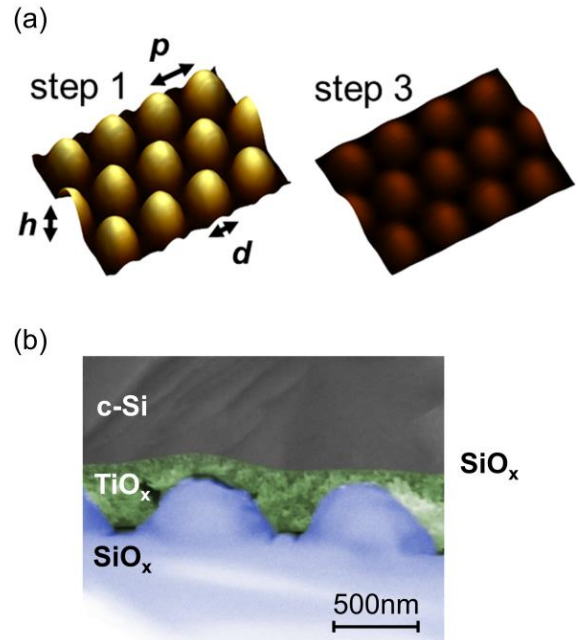

Fig. 2: (a) Exemplary atomic force microscope images showing the hexagonal $SiO_x$ nano-pillar array (cf. step 1 in Fig. 1) and the surface of the SMART texture (cf. step 3 in Fig. 1) with the same height scaling. The characteristic texture parameters are outlined. (b) Exemplary scanning electron microscope image of a SMART texture with a silicon absorber on top. For clarification, the $SiO_x$ (blue) and $TiO_x$ (green) layers have been colored. The thin $SiO_x$ passivation layer is not visible.

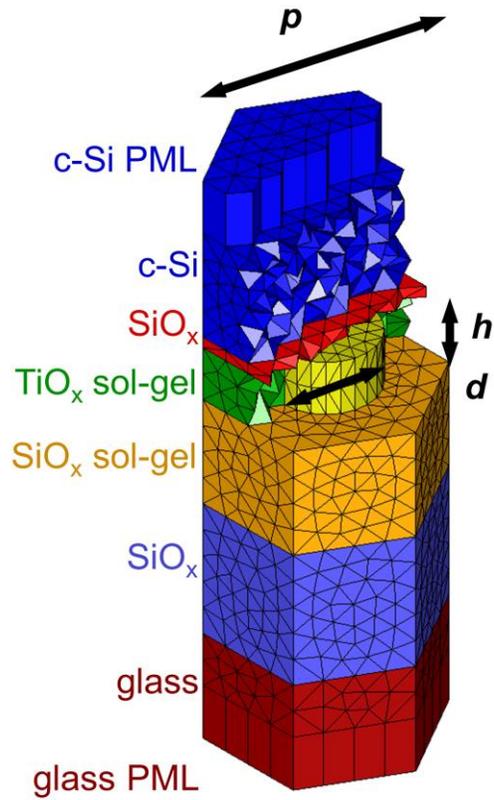

Fig. 3: Unit cell of the SMART texture used for 3-dimensional optical simulations, consisting of a hexagonal array of $SiO_x$ nano-pillars and the smoothing $TiO_x$ layer. The parameters varied in the simulations are denoted, namely the period ($p$), the height ($h$) and the diameter ($d$). On bottom and top, infinite halfspaces of glass and silicon layers are assumed, which is realized with perfectly matched layers (PML).

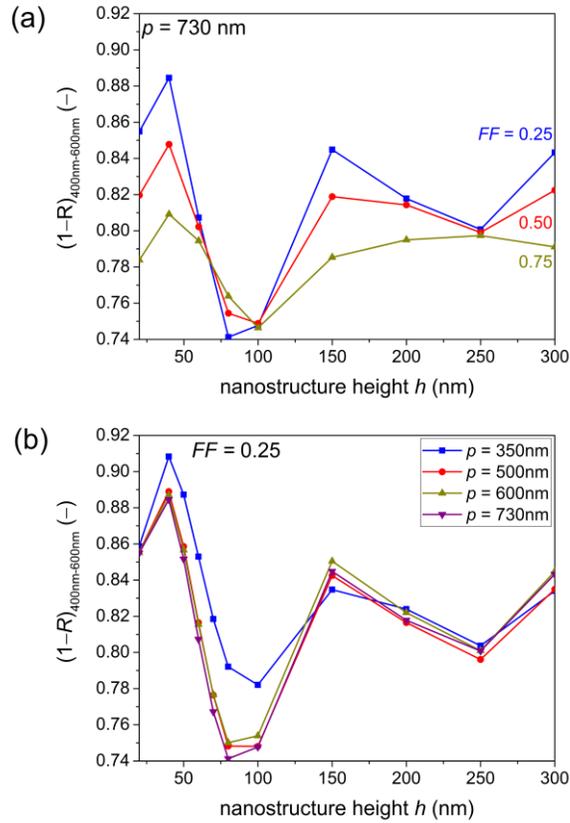

Fig. 4: Mean 1-R (reflectance) between 400 nm and 600 nm, calculated with 3-dimensional FEM simulations for (a) different filling fractions with a fixed period of 730 nm, and (b) various periods of the nanostructure with a fixed filling fraction of 0.25.

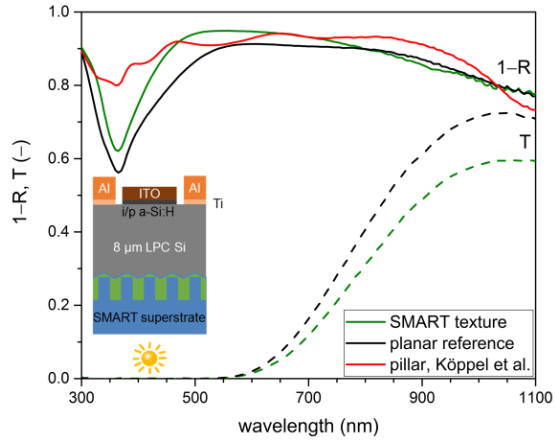

Fig. 5: Reflectance (represented as 1-R, solid lines) and transmittance (T, dashed lines) characteristics of liquid-phase crystallized silicon thin-film solar cells with 8 µm absorber layers comprising a SMART textured (green), planar reference (red) and nano-pillar array superstrate (red, data from Ref. (10)). The inset depicts a schematic of the solar cell stack on a SMART superstrate, with direction of the incident light indicated.

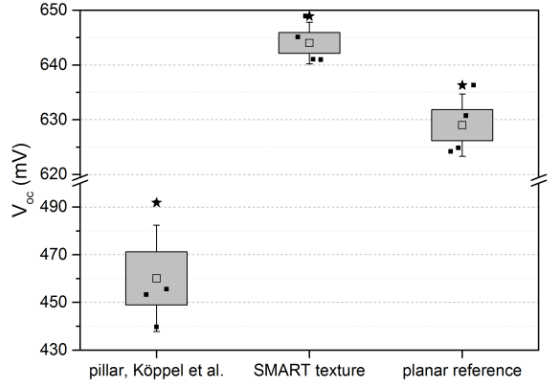

Fig. 6: Open-circuit voltage $V_{oc}$ of liquid-phase crystallized silicon thin-film solar cell stacks (see inset Fig. 5) measured by the Suns-$V_{oc}$ method illuminated from the glass side for the planar reference sample, the SMART texture and the hexagonal nano-pillar array published in Ref. [10]. Open squares represent the mean value of the best four cells (solid squares), boxes the standard error, and whiskers the standard deviation. The best measured $V_{oc}$ of each solar cell type is highlighted by a star.

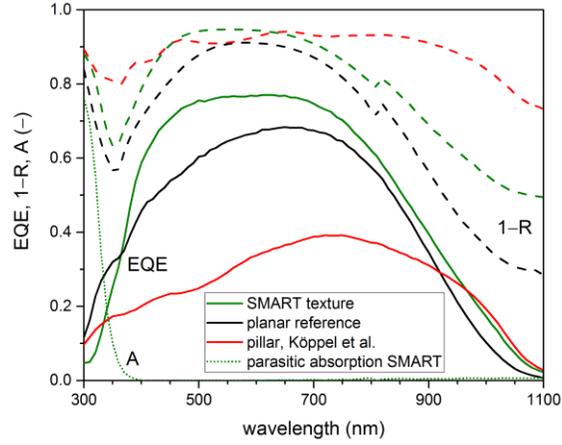

Fig. 7: Optoelectronic characterization of the SMART texture (green), planar reference (black) and nano-pillar array (red), as measured by external quantum efficiency (EQE, solid) and 1-reflectance (1-R, dashed). The dotted curve shows the parasitic absorption in the SMART texture (A, dotted).